# Smart Spectrometer for Distributed Fuzzy Control

E. Benoit, L. Foulloy

LAMII/CESALP, B.P. 806, F-74016 Annecy cedex, France

benoit@esia.univ-savoie.fr, foulloy@esia.univ-savoie.fr

## Abstract

If the main use of colour measurement is the metrology, it is now possible to find industrial control applications which uses this information. Using colour in process control leads to specific problems where human perception has to be replaced by colour sensors. This paper relies on the fuzzy representation of colours that can be taken into account by fuzzy controllers. If smart sensors already include intelligent functionalities like signal processing, or configuration, only few of them include functionalities to elaborate the fuzzy representation of measurements. In this paper, we develop a solution where the numeric processing is performed locally by the sensor, and where fuzzy processing is exported towards another computing resource by means of the CAN network. This paper presents the concept and the application to a smart fuzzy spectrometer.

Keywords: Intelligent sensor, spectrometer, fuzzy control, colour measurement

## Introduction

Colour is an important information in many industrial processes as for example food industry, or ink manufacturing, or cosmetics industry.

It can be a precise objective when manufacturing ink, paints, wallpaper... . But it is more imprecise when it is used for sorting fruits, or cooking biscuit, or for analysing the result of a chemical process. In this cases, decisions or control actions are based on human expertise.

Fuzzy sensors [1][2] were introduced to reproduce human-like perception. Therefore, the fuzzy representation of colour measurement seems to be an interesting solution for the integration of colour perception into control loop. Moreover, it provides a natural interface with a fuzzy rule based control of the process.

Obviously, classical sensors are not designed to provide a fuzzy representation of their measurement nor to perform fuzzy processing. This paper shows how this problem can be overcome using the concept of processorless intelligent sensor [3]. In section 1, the colour representation is addressed. Section 2 is devoted to the fuzzy representation of colours. In section 3 the concept of processorless smart sensors in a distributed environment is proposed. Finally, the application to a fuzzy spectrometer is discussed.

## 1. Colour Measurement

The colour sensing is scientifically studied since the famous prism experience performed by I. Newton. Nowadays, it is well-known that the information of colour associated with a light beam is included in the spectral energy distribution of the electromagnetic wave flow of this light beam. The light beam can be issued from an illuminant source, or be issued from an illuminated object.

The CIE (International Commission on Illumination) defines several recommendations for the colour measurement of surfaces [4]. The aim of these propositions is to normalise the artificial colour perception and the colour representation. As most users need to copy human perception, CIE recommendations allow a human-like colour perception.

### 1.1 Basic Measurement

One of these recommendations is to perform measurement of light with 3 photo-sensors. The spectral sensitivity of these sensors needs to be respectively one of the colour-matching functions $\bar{x}(\lambda)$, $\bar{y}(\lambda)$ and $\bar{z}(\lambda)$ (see Fig. 1). Measures, i.e. the tristimuli X Y and Z, can be also obtained with a spectrometer by integration over the spectrum.

For example, X is given by Eq. (1) where $\bar{x}(\lambda)$ is a colour-matching function, $\phi(\lambda)$ is the spectral energy distribution acquired by the spectrometer, and $k$ is a constant.

$$X = k \cdot \int_{380nm}^{780nm} (\phi(\lambda) \cdot \bar{x}(\lambda))d\lambda \qquad (1)$$

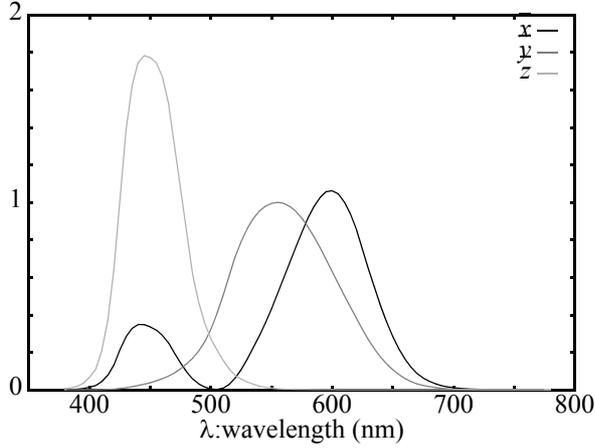

Fig. 1 colour-matching functions $\bar{x}(\lambda), \bar{y}(\lambda), \bar{z}(\lambda)$

The last step is the calculation of chromaticity coordinates $x, y$ and $z$. These coordinates are defined in order to separate the illumination information and the chromatic information. A coordinate is calculated by dividing a tristimulus by the sum of the three tristimuli:

$$x = \frac{X}{X + Y + Z} \qquad (2)$$

Because of the relation $x+y+z=1$, only two coordinates are used to define a colour. Commonly, $x$ and $y$ coordinates are used and defined the so-called $xy$-plane.

## 1.2 Colour Spaces

The $xyz$ colour space is not very close to human perception. Several experiments have been performed to make the comparison between a distance of two colours into the $xy$-plane and the subjective distance of the same colours. The result of these experiments shows that the $xy$-plane is not uniform in the context of a human perception, i.e. there is large difference between these distances. In 1976, the CIE proposes several approximately uniform colour spaces in order to reduce this difference.

For control applications based on human-like perception, we chose the Lab colour space. In this colour space, the distance operator can be considered as uniform, i.e representative of a subjective colour space. This space was chosen because the knowledge about the role of colour into a control application relies on a human expertise. This remark induces the choice of a colour representation close to human one.

$$L = 116 f\left(\frac{Y}{Y_n}\right) - 16 \qquad (3)$$

$$a = 500\left(f\left(\frac{X}{X_n}\right) - f\left(\frac{Y}{Y_n}\right)\right) \qquad (4)$$

$$b = 200\left(f\left(\frac{Y}{Y_n}\right) - f\left(\frac{Z}{Z_n}\right)\right) \qquad (5)$$

with

$$\left(f(d) = d^{\frac{1}{3}}\right) if (d > 0.008856) \qquad (6)$$

$$\left(f(d) = 7.787(d) + \frac{16}{116}\right) else \qquad (7)$$

On the Fig. 2, common colours are represented on the Lab colour space.

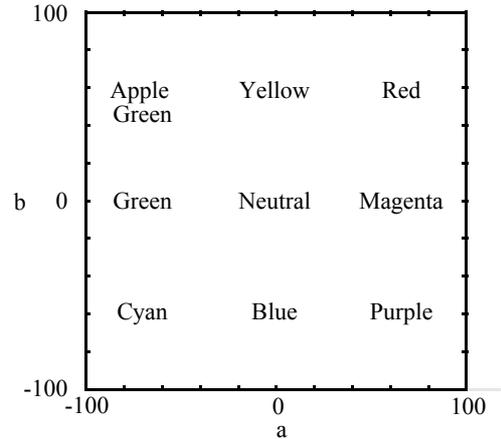

Fig. 2 Lab colour space

## 2. Fuzzy Representation

If numerical values are commonly used to represent measurement results, it is now admitted that some applications would better handle symbolic values or linguistic terms. In this case, a measurement result is represented by a fuzzy subset of linguistic terms. The conversion from numerical to linguistic representation is called a linguistic description [5].

## 2.1 Fuzzy Meaning and Description

To perform a linguistic measurement, it is necessary to clearly specify the relation between linguistic terms and numbers [2].

Let $X$ be the universe of discourse associated with the measurement of a particular physical quantity. Denote $x$ any element of $X$. In order to linguistically characterise any measurement over $X$, let $W$ be a set of words, representative of the physical phenomenon. For example, the set $W=\{$*Dark*, *Medium*, *Bright*$\}$ can be used to represent an illumination. Denote $F(E)$ the set of the fuzzy subsets of a set $E$.

Introduce an injective mapping $M : W \rightarrow F(X)$, called the *fuzzy meaning* of a linguistic terms (see Zadeh [5]). It associates any term $L$ of $W$ with a fuzzy subset of $X$. Injectivity guarantees that two different terms may not have the same meaning. In other words, two synonymous terms have the same meaning. The fuzzy meaning of a symbol $L$ is characterized, for all $x \in X$, by its membership function denoted $\mu_{M(L)}(x)$.

Another mapping $D : X \rightarrow F(W)$, called the *fuzzy description* of a measurement over $W$ associates any measurement of $X$ with a fuzzy subset of linguistic terms of $W$. The fuzzy description of a measurement is characterized, for all $L \in W(X)$, by its membership function $\mu_{D(x)}(L)$. Any measurement that belongs to the meaning of a linguistic term, can obviously be symbolically described at least by this term. Therefore, the following relation links the description of a measurement to the meaning of a term:

$$\mu_{D(x)}(L) = \mu_{M(L)}(x) \qquad (8)$$

It means that if a linguistic terms belongs to the description of a measurement at a grade of membership $\mu_{D(x)}(L)$, then the measurement belongs to the meaning of the linguistic terms at the same grade of membership. Let $X = [0, 100]$ be the universe of discourse for the measurement of illuminations. The fuzzy meanings of *Dark*, *Medium*, and *bright* are represented by the membership functions (see Fig. 1). The fuzzy description of the measurement $x = 80\%$ is obtained according to Eq. (8) and is equal to $\{0.25/$*Medium*, $0.75/$*Bright*$\}$[1]. This means that the fuzzy description of an illumination of 80% is a fuzzy subset which contains the linguistic term *Bright* with a grade of membership of 0.75 and the linguistic term *Medium* with a grade of membership of 0.25. This fuzzy description does not contains the term *Dark*.

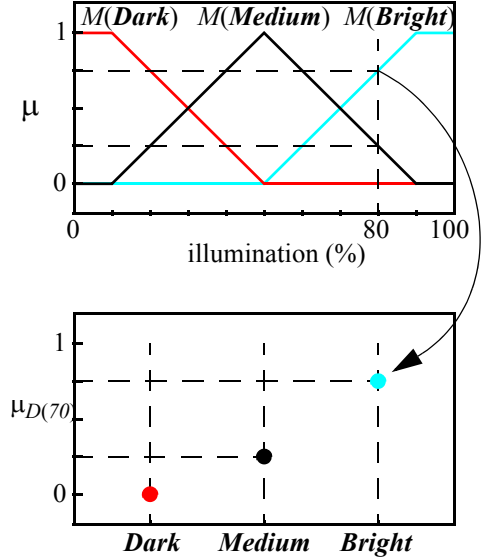

Fig. 3 Fuzzy meanings and fuzzy description

## 2.2 Fuzzy Representation of Colours

From what preceeds, a fuzzy representation of a colour is a fuzzy subset like $\{0.52/$*Red*, $0.3/$*Neutral*, $0.18/$*Yellow*$\}$. Two ways can be chosen to build a fuzzy representation of colour.

The first one is based on the direct description of a chromatic coordinate [1]. It extends the concept of fuzzy meanings from 1 to 3 dimensions and requires a learning phase based on a set of examples.

The second one is based on a separate fuzzy description of each colour coordinate associated with a rule-based aggregation. This one is preferred when the space has a regular cubic structure like in the Lab space. Indeed the Lab space can be cut into nine squares (Fig. 2).

A fuzzy description is extracted from each coordinate $a$ and $b$. We use a simple linguistic set for both coordinates: $L=\{$*Negative*, *Null*, *Positive*$\}$. The meanings of the linguistic terms are represented in Fig. 4

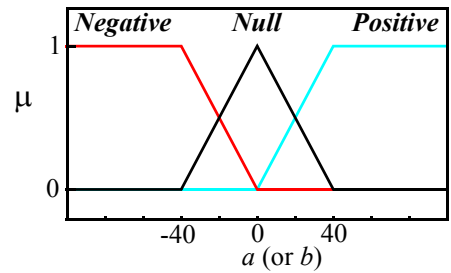

Fig. 4 Fuzzy meaning of *Negative*, *Null* and *Positive*

---

1. Terms with a null grade of membership are omitted.

Fuzzy meanings are chosen regarding to the subjective perception of colours in Lab space. We consider that the saturation of a colour is close to the maximum when the coordinate *a* or *b* is over 40 or below -40. So the fuzzy meaning of the term **Positive** includes all values over 40 and includes partially and progressively values from 0 to 40. The same approach is used to define the fuzzy meaning of **Negative**. The fuzzy meaning of the symbol **Null** is deduced by applying the constraint of fuzzy partition.

Now, a fuzzy description of colour with usual linguistic terms like **Red** or **Yellow** is obtained from a human expertise expressed with a set of rules, and from the values of *a* and *b*.

if *a* is **Positive** and *b* is **Positive** then *colour* is **Red**
if *a* is **Null** and *b* is **Positive** then *colour* is **Yellow**
...

The linguistic set for colours is {**Red**, **Green**, **Blue**, **Yellow**, **Cyan**, **Magenta**, **Purple, AppleGreen**, **Neutral**}. The following rule-base is used to perform the fuzzy aggregation

| colour | | *a* | | |
|---|---|---|---|---|
| | | **Negative** | **Null** | **Positive** |
| *b* | **Negative** | Cyan | Blue | Purple |
| | **Null** | Green | Neutral | Magenta |
| | **Positive** | AppleGreen | Yellow | Red |

Fig. 5 Set of rules for fuzzy colour representation

This set of rules is simple but it is can be extended by using more symbols for the fuzzy representation of *a* and *b*. For example, with the set of symbols L={**VeryNegative**, **Negative**, **Null**, **Positive, VeryPositive**} a set of 25 rules can be defined. Another method consists in automatic building of rule set.

## 2.3 Fuzzy control

The fuzzy control is based on the same principles than fuzzy aggregation. It uses on one or more symbolic inputs and produces a symbolic output. The difference is that the result of a fuzzy control can be expressed in a numerical format. In addition to a rule based processing, a fuzzy controller is able to perform a symbolic to numeric conversion, the so-called defuzzification.

In the case of a process driven with a colour information, one of the entries is a fuzzy description of colour.

# 3. Distributed Control

## 3.1 Smart Sensor Concept

Since the eighties, it is commonly admitted that smart sensors rely on functionalities such as measurement, communication, configuration and validation [6]. Even if measurement remains the most important function of a sensor, other functionalities are become increasingly used into present intelligent sensors.

## 3.2 Distributed Intelligence

Nowadays, more and more functionalities are included into sensors. This induces to include complex computing resources into these kind of sensors. In precedent papers [7][3], we proposed to code the definition of these functionalities in a specific language (PLICAS), and to implement them on the smart sensor even if this one do not have enough computing resource to perform these functionalities. In that case, the functionalities coded in PLICAS are sent over a network to be executed by another computing system. This approach allows to add complex functionalities to usual intelligent instruments.

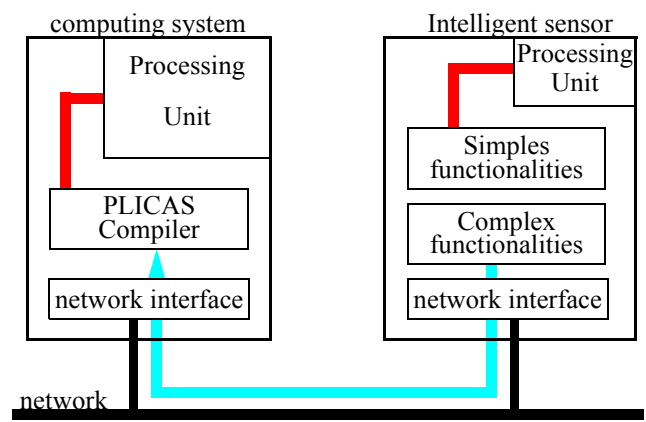

Fig. 6 Concept of intelligent sensors with small processor

The PLICAS language owns basic functionalities of sequential programming. In order to minimise the size of the compiler code, the language uses only global variables and does not allow new function definitions. The language was created to manage the sequencement of predefined functions. It includes basic arithmetic and logical functions needed to perform this sequencement. The PLICAS language was build in order to allow a quick conception of fuzzy sensors.

## 4. Application

### 4.1 Hardware

The spectrometer that is used is a Zeiss monolithic miniature spectrometer with a spectral range from 300nm to 900nm. This sensor is implemented into a board based on the 80C196KC micro-controller. This board performs basic measurements and signal processing. It includes services of spectral measurement, black level compensation, white correction, XYZ calculation, Lab calculation, and communication.

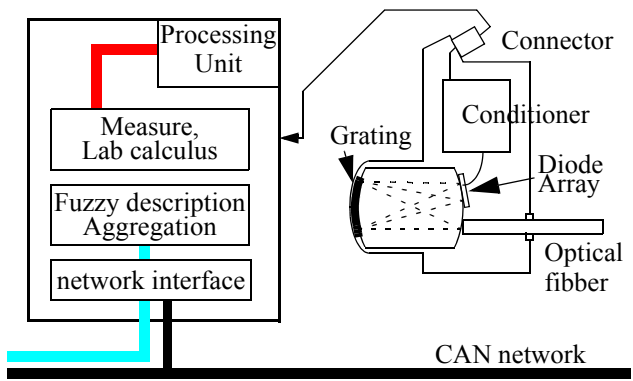

Fig. 7 The fuzzy spectrometer

The micro-controller board includes two large permanent memories (ReadOnlyMemory). One contains the executive code for the management of the sensor, the Lab calculus and the other measurement services. The other one contains the PLICAS source code of the fuzzy description and the fuzzy aggregation.

The board includes a network interface in order to communicate with other sensors or computers through the CAN network (Controller Area Network). The CAN network is preferentially used in automotive industry and manufacturing industry. It is dedicated to the communication of small data (8 bytes maxi) like sensor results. A specific communication protocol has been implemented in order to exchange PLICAS source code through the CAN network.

The PLICAS compiler is implemented on a PC with a CAN network interface. The PC acts as a computing resource server. It catch all PLICAS source codes sent on the network, then it compiles the source code and executes it. The communication of consumed values or produced values have to be defined in the code.

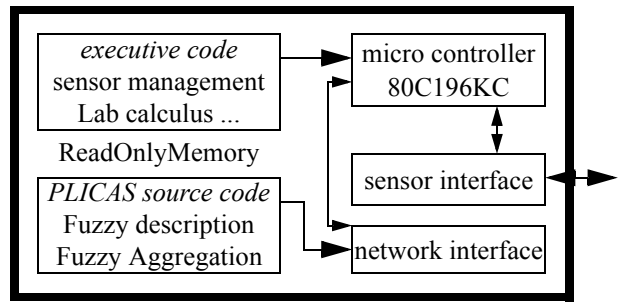

Fig. 8 Description of the micro-controller board

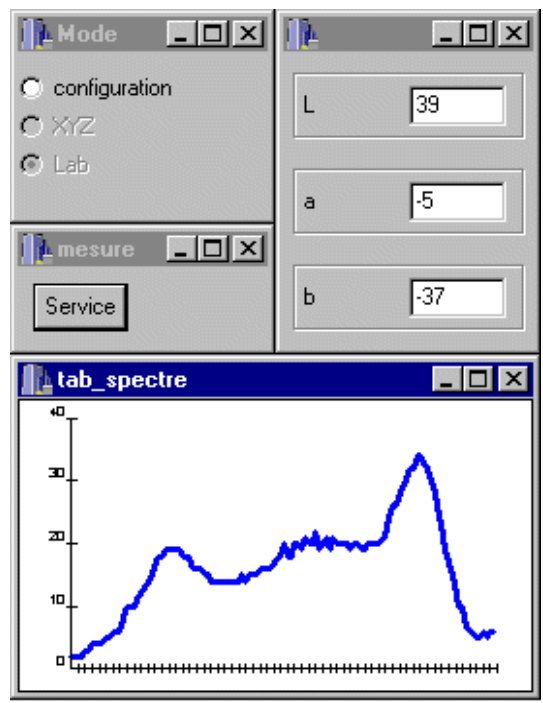

Fig. 9 Values produced by the smart spectrometer

### 4.2 Conception

The signal processing is cut into two parts. The numeric signal processing is implemented on the spectrometer. The definition of the fuzzy signal processing is written into memory.

Concerning the numeric processing, the sensor is a quite classical smart sensor. Services defined are:
- spectral measurement
- black level compensation
- white correction
- sensitivity adjustment

- XYZ calculation
- Lab calculation
- CAN network communication

The PLICAS code provided in the annex performs the fuzzy description of *a* and *b* coordinates and performs an aggregation in order to obtain the fuzzy description of colour. The code is cut into 4 blocks.

The "declarations" block defines all variables. Variables *L*, *a* and *b* are defined as *external* which means that they are collected on the network. Variables *pab*, and *pcolour* are sets of fuzzy meanings. Variables *fa* and *fb* are fuzzy descriptions, i.e. fuzzy subsets of linguistic terms. Variable *fcol* defined as *public* is also a fuzzy description but it can be sent over the network.

The "initialisation" block includes a list of instructions. This block is executed only one time, just before the "main" block. In this block, we find the definition of a set of meanings. The function *def_partition*() creates a new set of meanings *pab* and initialises the meaning of the term ***Null***. Functions *above()* and *below()* define meanings of terms ***Neg*** and ***Pos*** relatively to the meaning of the term ***Null***. Links between this set and fuzzy representations of *a* and *b* are also defined (*fa in pab*).

The "main" block includes also a list of instructions. It is executed periodically. Functions *import*() and *export*() are used for network exchanges. The function *fuzz*() perform a fuzzy description. The function *execute*() executes the last block.

The "inference" block includes the set of rules written in a simple format.

## 4.3 Results

The following example shows the result obtained with a green surface. When the sensor starts, it sends its PLICAS code throw the network. After acquiring the spectral density, the smart spectrometer calculates the Lab coordinates:

a = -37.97, b=9.85    (9)

These coordinates are sent over the network and caught by the PC which runs the fuzzy description. Fuzzy descriptions of *a* and *b* are then:

*D*(a) = {0.941/Negative, 0.059/Null}    (10)

*D*(b) = {0.745/Null, 0.255/Positive}    (11)

Then, the fuzzy representation of the colour is calculated:

{0.043/Neutral, 0.698/Green, 0.012/Yellow, 0.239/AppleGreen}    (12)

# Conclusion

Sometimes, measurements are so close to subjective human perception that it seems very difficult to use them to control a process. The idea presented in this paper is to use, both for both measurement and control processing, a representation close to the human one.

Fuzzy techniques are frequently used in industry especially for process control, but are not enough developed for sensors. The concept presented in this paper can be a solution to implement complex fuzzy functionalities or very specific functions in sensors by mean of a dedicated language. The source code of these complex functionalities is then exported over the network in order to be executed by a remote system. For example dedicated units like fuzzy processors for fuzzy calculus or DSP (Digital Signal Processing) for complex and high-speed calculus can be used.

# Annex: PLICAS source code

```
declarations
    external double L,a,b;
    partition pab, pcolour;
    subset fa,fb,fcol;
    public subset fcol;
block initialisation
    def_partition(pab,"Null",-40.0, 0.0, 0.0, 40.0,-100.0, 100.0);
    above(pab,"Null","Pos");
    below(pab,"Null","Neg");
    def_symbols(pcolour, 9, "Neutral", "Red", "Green", "Blue", "Cyan", "Magenta", "Yellow", "AppleGreen", "Purple");
    fa in pab;
    fb in pab;
    fcol in pcolour;

block inference
    fcol=0.0;
    if fa is "Null" and fb is "Null" then fcol is "Neutral";
    if fa is "Null" and fb is "Pos" then fcol is "Yellow";
    if fa is "Null" and fb is "Neg" then fcol is "Blue";
    if fa is "Pos" and fb is "Null" then fcol is "Magenta";
    if fa is "Pos" and fb is "Pos" then fcol is "Red";
    if fa is "Pos" and fb is "Neg" then fcol is "Purple";
    if fa is "Neg" and fb is "Null" then fcol is "Green";
    if fa is "Neg" and fb is "Pos" then fcol is"AppleGreen";
```

```
    if fa is "Neg" and fb is "Neg" then fcol is "Cyan";
block main
    import(L);
    import(a);
    import(b);
    fa = fuzz(a);
    fb = fuzz(b);
    execute(inference);
    export(fcol);
```